# Giant Resonant Reflection Gain from Injection-Induced Quenching in a Tunnel Diode

Darmindra Arumugam*, Jack Bush, Brook Feyissa
Jet Propulsion Laboratory, California Institute of Technology, Pasadena, 91109, California, USA

*Corresponding author. E-mail(s): darmindra.d.arumugam@jpl.nasa.gov;

**Abstract**
Negative-resistance microwave oscillators can simultaneously sustain autonomous oscillations and coherently scatter RF signals, enabling active reflection beyond conventional linear amplification. Here we demonstrate large resonant reflection gain from synchronization-induced phase localization in a self-sustained tunnel-diode oscillator operating near 2.96 GHz. Weak resonant RF injection drives a transition from a broadband free-running state to a phase-localized narrowband reflected carrier. The resulting coherent reflected carrier reaches $83.1 \pm 0.8$dB relative to the injected signal and is accompanied by order-of-magnitude linewidth collapse, nonlinear injection pulling, and a synchronization bandwidth of 15.84 kHz. Time-resolved spectrograms directly highlight injection locking and frequency entrainment, while noisy Stuart–Landau simulations reproduce the observed spectral concentration and detuning-dependent gain roll-off near the Hopf instability. Artificially increasing phase diffusion with broadband bias noise suppresses the reflected enhancement, supporting synchronization-induced quenching of phase fluctuations as the mechanism underlying the large coherent reflection gain.

Negative-resistance microwave devices can simultaneously sustain autonomous oscillations and amplify reflected electromagnetic waves, placing them at the boundary between oscillators and amplifiers[1,2]. Reflection amplifiers based on transistors, Gunn diodes, IMPATT diodes, and tunnel diodes have long been investigated for low-noise microwave amplification, active reflection, repeater architectures, and radar backscatter or communication systems[3-6]. In particular, modern germanium tunnel-diode reflection amplifiers have demonstrated compact low-power microwave operation with reflection gains exceeding 30dB in impedance-engineered negative-resistance configurations[7-8]. However, prior work primarily treated these devices as engineered active loads or reflection amplifiers, rather than explicitly exploiting synchronization and phase-diffusion quenching in a self-sustained oscillatory regime. Here, we directly resolve the transition from the broad free-running state to a narrow synchronized reflected carrier under weak injection, revealing a route to exceptionally large coherent reflection through synchronization of an autonomous oscillator.

Injection locking and phase synchronization are universal phenomena in nonlinear oscillatory systems and play a central role in coherent microwave sources, nonlinear wave dynamics, and coupled oscillator networks[9-11]. Near a Hopf instability, weak coherent forcing can strongly suppress stochastic phase diffusion, producing phase localization and concentration of oscillatory energy into a narrow synchronized spectral mode[12-14]. While these synchronization phenomena are well known theoretically, their role in active microwave reflection and negative-resistance scattering systems has remained largely unexplored experimentally.

Here we show that a tunnel-diode oscillator operated in a self-sustained regime can exhibit large resonant reflection gain through injection-induced quenching of phase diffusion. A weak resonant RF tone drives a transition from a broadband free-running oscillation to a synchronized state with a localized oscillator phase (‘phase-localized’ state) in which oscillator energy collapses into a narrow coherent reflected carrier. The resulting reflected enhancement exceeds 80 dB (83.1±0.8 dB) relative to the injected signal and is accompanied by linewidth collapse, injection pulling, finite locking bandwidth, and strong sensitivity to externally induced phase diffusion. A noisy Stuart–Landau description captures the observed detuning roll-off and linewidth quenching, linking the microwave response to synchronization dynamics near a noisy Hopf instability[9-11]. These results establish synchronization-enabled active reflection as a powerful route to exceptionally large coherent reflection enhancement in self-sustained nonlinear oscillators.

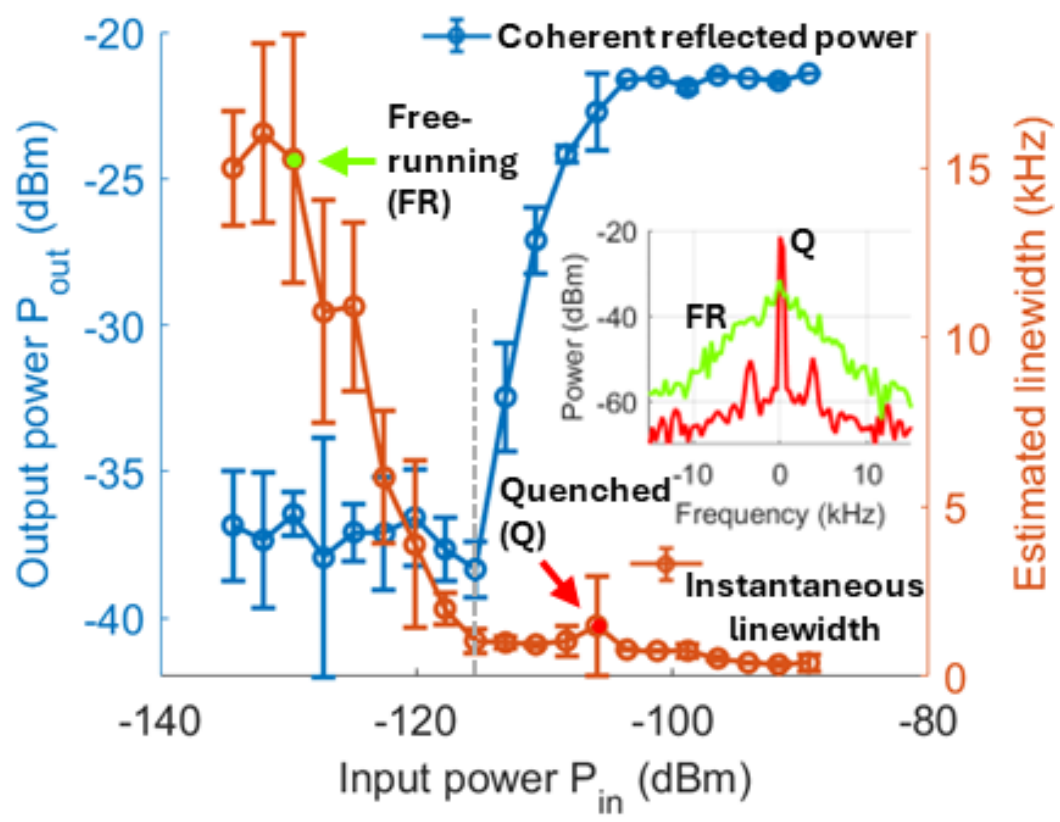

**Fig. 1: Injection-induced quenching concentrates oscillator power into a coherent reflected carrier:** Increasing resonant injection drives a transition from a free-running noisy oscillation to a phase-localized state in a tunnel-diode oscillator (TDO). As the injected power increases, the instantaneous linewidth (right axis) collapses while the coherent reflected power (left axis) rises sharply, revealing suppression of phase diffusion and concentration of oscillation energy into a synchronized spectral component. The dashed line marks the onset of quenching, where coherent ordering emerges and reflected power rapidly increases. The inset shows representative spectra illustrating the transformation from a broadband free-running state (FR) to a narrow-quenched state (Q).

## Results

Injection-induced quenching in a tunnel-diode oscillator (TDO) produces a simultaneous enhancement of coherent reflected power and collapse of the oscillator linewidth, providing a direct signature of phase localization. Measurements are performed near 2.96 GHz using a DC-biased TDO coupled through a circulator, in which a weak injected RF tone drives the oscillator and the reflected signal is monitored. As the injected RF power $P_{in}$ is increased, the system transitions from a free-running regime, characterized by broad spectral emission and weak reflection, to a phase-localized state in which a narrow, coherent carrier dominates the response (Fig. 1). This transition is marked by a rapid rise in reflected power alongside an order-of-magnitude reduction in instantaneous linewidth, indicating suppression of phase diffusion and the emergence of phase entrainment. The dashed line denotes the onset of this quenching regime, where coherent dynamics begin to dominate over stochastic fluctuations. Reflection gain is defined as $10\log_{10}(P_{\text{out}}/P_{\text{in}})$ (in dB). Representative spectra (inset of Fig. 1) illustrate the corresponding spectral transformation, showing the collapse of a broadband pedestal into a narrow, synchronized peak. These observations establish that resonant injection can concentrate oscillator energy into a coherent reflected signal, yielding large effective gain through synchronization-induced phase stabilization rather than conventional amplification.

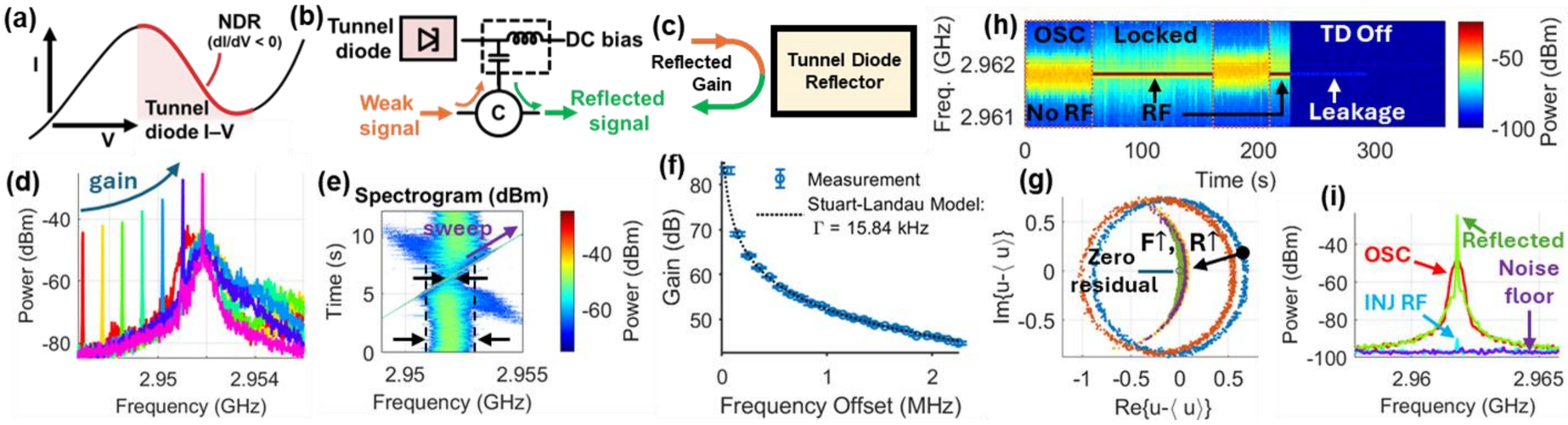

**Fig. 2: On-resonant reflection gain in a tunnel-diode oscillator (TDO).** (a) Tunnel-diode I–V characteristic highlighting the negative differential resistance (NDR, dI/dV < 0) region that provides negative damping and sustains a nonlinear microwave limit cycle. (b) Experimental configuration: a DC-biased tunnel diode (TD) coupled via a circulator; a weak injected RF tone drives the TDO and the reflected port contains both injected and oscillator fields. (c) Nonlinear reflector concept: the self-oscillatory (OSC) diode acts as an active mirror whose reflection is governed by Hopf-type amplitude–phase dynamics. (d) Measured reflection spectra near 2.962 GHz for varying injection detuning frequency, showing progressive enhancement of the coherent reflected carrier as the detuning approaches zero. (e) Time–frequency spectrogram during a frequency sweep across OSC resonance; as detuning approaches zero, the OSC linewidth collapses and power concentrates into the synchronized component (quenching). (f) Reflection gain versus detuning; data follow a Stuart–Landau roll-off with locking bandwidth Γ = 15.84 kHz, consistent with injection-modified Hopf dynamics. (g) Residual phase portraits in the rotating frame $(u - \langle u \rangle)$; F is normalized injection amplitude and $R = |\langle e^{i\phi} \rangle|$ the phase-coherence order parameter. Increasing F (0→0.5) drives R≈0.15→0.99 and shrinks the residual cloud toward the coherent fixed point (origin). (h) Measured spectrogram identifying the free-running oscillator (OSC), injection-locked (Locked), and tunnel-diode-off (TD Off) conditions, with the TD Off condition providing the direct injected-carrier/feedthrough baseline.. (i) Representative reflection spectrum under strong injection showing OSC and reflected signal relative to injected RF (INJ) and system noise floor.

The physical origin of the quenching-induced gain is illustrated in Fig. 2, which connects the device-level nonlinearity to the observed phase-coherent response. The tunnel-diode operates in a negative differential resistance (NDR) regime (Fig. 2a), providing the negative damping required to sustain a nonlinear limit cycle. When coupled through a circulator (Fig. 2b), the oscillator acts as an active reflector: a weak injected RF tone interacts with the self-sustained oscillation, and the reflected field contains both injected and oscillator-driven components. This interaction can be understood as a nonlinear scattering process governed by amplitude–phase dynamics near a Hopf bifurcation (see Theoretical Framework), in which the limit-cycle oscillator functions as a gain-enabled reflector (Fig. 2c). In this picture, the injected field synchronizes the oscillatory phase, and the resulting phase-coherent response acts to redirect energy from the self-sustained oscillation into the reflected signal, enabling reflected gain. The oscillator therefore behaves as a driven nonlinear active reflector whose effective reflection coefficient is dynamically modified by phase entrainment. (see Theoretical Framework).

The measured response under frequency detuning directly reveals this nonlinear interaction. As the injected frequency approaches the free-running oscillator resonance, the reflected spectrum exhibits a pronounced increase in amplitude (Fig. 2d), accompanied by a collapse of spectral width in time-resolved spectral-sweep of injected RF measurements (Fig. 2e). This behavior indicates the onset of synchronization, where phase fluctuations are suppressed and oscillator energy is concentrated into a coherent spectral component. The increase in coherent carrier power reflects synchronization-mediated concentration of RF power already sustained by the DC-driven autonomous oscillator and does not require a corresponding increase in total integrated RF output power. The resulting reflection gain exhibits a sharp maximum at zero detuning (Fig. 2f), consistent with injection-locking dynamics and well captured by a Stuart–Landau description of the oscillator response (see Theoretical Framework).

The corresponding model phase-space representation further illustrates this transition. In the rotating frame, the oscillator phase evolves from a diffusive distribution to a localized fixed point as the injection strength increases (Fig. 2g), with the phase-coherence parameter approaching unity. This phase localization directly corresponds to the emergence of a narrow coherent carrier in the reflected signal. Experimental spectra (Fig. 2h,i) provide direct experimental evidence for this transition, showing the evolution from free-running oscillation (OSC) to injection-locked behavior and, under strong drive, a dominant reflected component exceeding the injected signal. Together, these observations support the interpretation that the large on-resonant reflection gain arises from injection-induced phase ordering of the nonlinear oscillator, rather than conventional linear amplification.

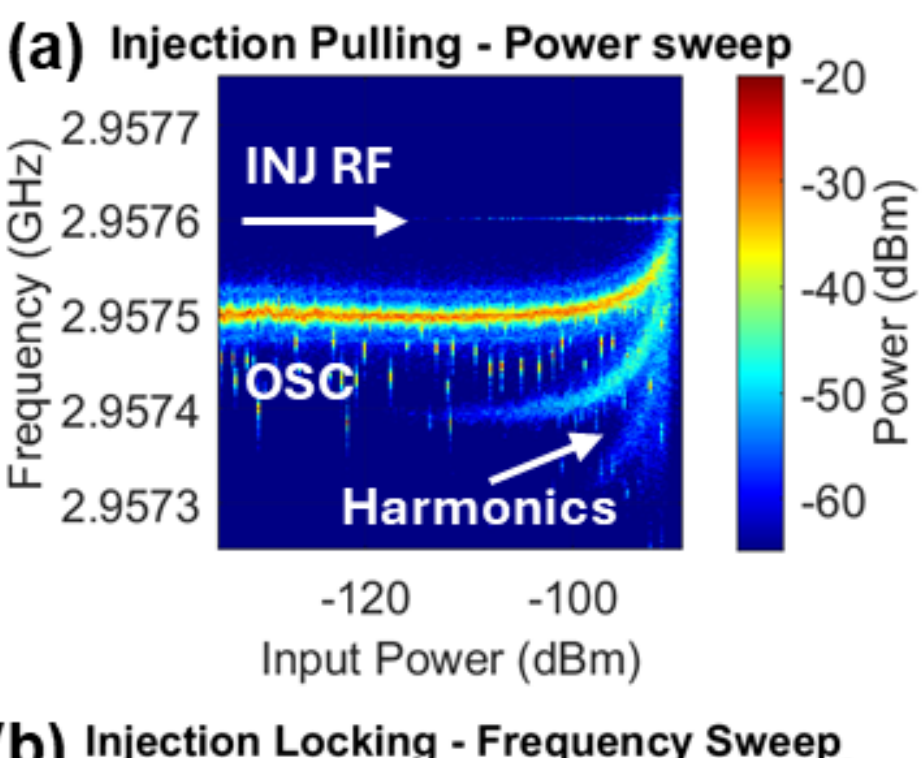

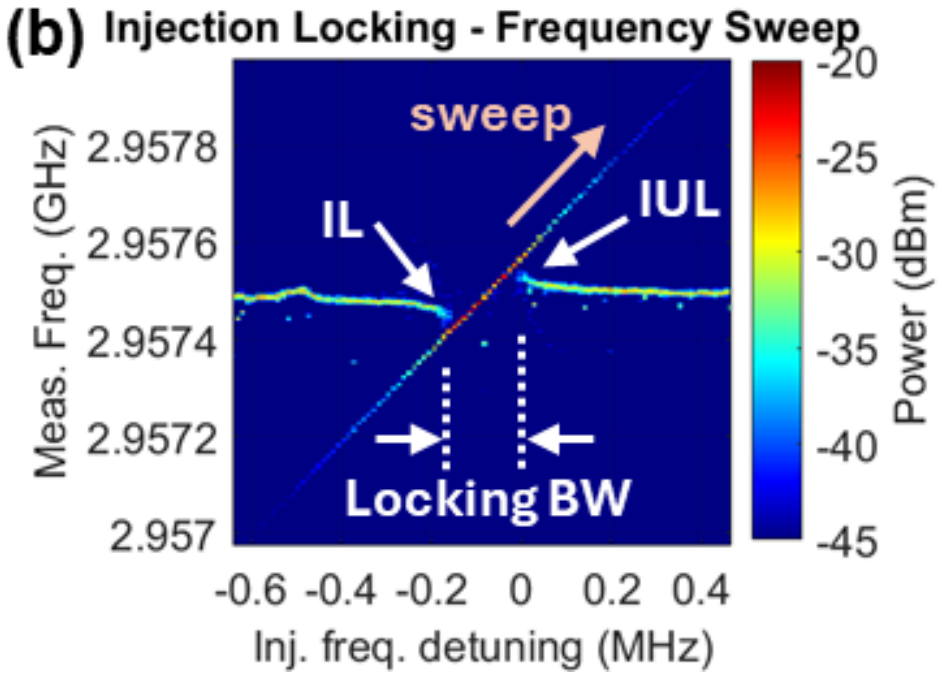

**Fig. 3: Injection pulling and locking dynamics of the tunnel-diode oscillator.** (a) Injection pulling (power sweep). Measured reflection spectrogram near 2.9GHz while sweeping injected RF power from −134dBm to −89dBm. The horizontal branch corresponds to the free-running self-sustained oscillation (OSC). Increasing injection power pulls the oscillator frequency toward the injected tone (INJ RF) through nonlinear amplitude–phase coupling of the NDR-driven limit cycle. The strong curvature at highest powers marks approach to synchronization. (b) Injection locking (frequency sweep). Reflection spectrogram versus injected-frequency detuning (−0.4 to +0.6 MHz) at fixed injection amplitude. The diagonal trace denotes the swept INJ RF. Within a finite detuning interval (dashed lines), the oscillator is frequency-entrained (IL), collapsing onto the drive. Outside this interval, injection unlocking (IUL) occurs and the oscillator reverts to detuned free-running behavior. The locking bandwidth is defined by the detuning span over which entrainment persists, consistent with Adler/Stuart–Landau phase dynamics.

To establish that the large reflected gain originates from synchronization rather than conventional amplification, it is necessary to resolve how the nonlinear oscillator evolves under increasing injected drive and finite frequency detuning. As the injected RF power increases (Fig. 3a), the free-running self-sustained oscillation is progressively pulled toward the injected carrier through nonlinear amplitude–phase coupling, producing harmonic structure characteristic of the strongly non-isochronous tunnel-diode dynamics. Near the quenching threshold, the oscillator abruptly collapses into a phase-localized state in which the spectral response concentrates onto the injected tone, suppressing phase diffusion and enabling coherent reflected enhancement. Under finite detuning (Fig. 3b), this synchronized state persists only within a bounded locking interval, outside of which the oscillator reverts to an injection-unlocked regime. The resonant reflection gain therefore emerges specifically within the injection-locked phase of the nonlinear oscillator, demonstrating that the reflected enhancement is governed by synchronization-induced phase ordering near the Hopf instability rather than linear small-signal amplification. The observed pulling and locking behavior is consistent with established Adler-type injection synchronization and establishes the dynamical basis for the coherent reflection enhancement observed in the synchronized state.

To identify the dynamical origin of the synchronization-induced gain, numerical simulations based on a noisy Stuart–Landau (Hopf) oscillator with coherent RF forcing and phase-diffusive noise were performed. The ensemble-averaged spectra in Fig. 4a reproduce the transition from a broadband free-running state to a spectrally concentrated synchronized response as the injected tone approaches resonance. With decreasing detuning, spectral weight progressively collapses into the injected component while the incoherent pedestal is suppressed, indicating injection-induced phase localization of the nonlinear limit cycle. Focusing on the near-resonant of the simulated response (Fig. 4b) shows the emergence of a sharp coherent carrier accompanied by strong suppression of the broadband phase-diffusive background, directly revealing quenching of stochastic phase fluctuations in the synchronized state.

The simulations closely reproduce the experimentally measured spectral evolution under weak resonant injection (Fig. 4c), including both the linewidth collapse and formation of a dominant coherent reflected peak. The coherent enhancement rapidly diminishes with increasing detuning, producing the monotonic gain roll-off shown in Fig. 4d. Simulated and measured gain curves exhibit strong agreement over the locking region and are well described by noisy Adler/Stuart–Landau synchronization dynamics near a Hopf instability (Theoretical Framework), yielding an extracted locking bandwidth of $\Gamma = 15.84$kHz. These results support the interpretation that the large coherent reflected enhancement arises from injection-induced phase localization and coherent spectral concentration of stored oscillation energy in the reduced synchronization model, rather than conventional linear amplification.

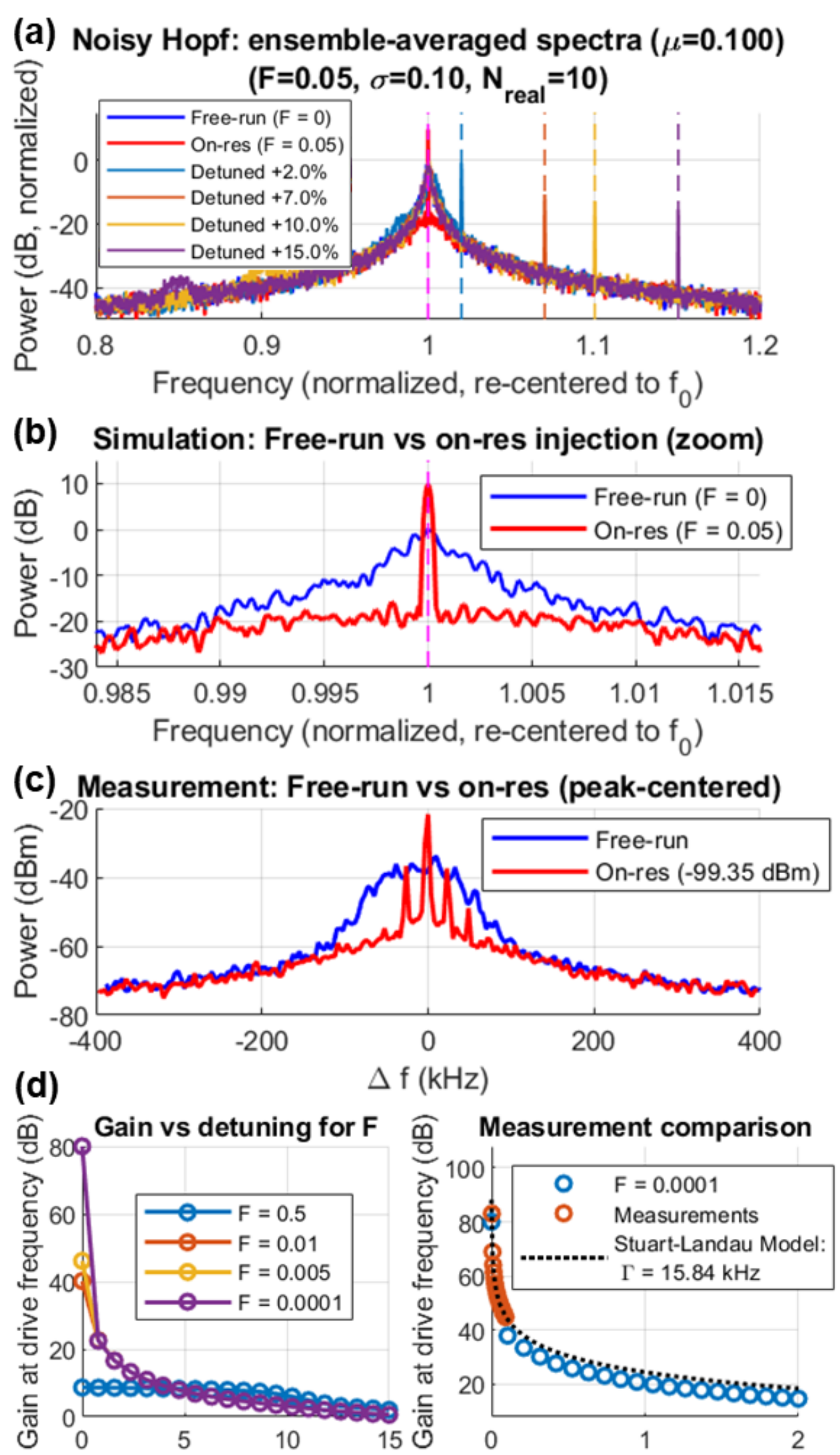

**Fig. 4: Noisy Hopf model of injection-enhanced reflection and detuning-dependent gain.** (a) Ensemble-averaged power spectra of a noisy Hopf oscillator (μ = 0.10, σ = 0.10, F = 0.05), normalized to the free-running peak and re-centered to the measured natural frequency $f_0$. Free-run (F = 0), on-resonance injection, and positive detunings (+2–15%) are shown. As detuning decreases, spectral weight collapses into the injected component and the linewidth narrows, evidencing phase entrainment and quenching of incoherent fluctuations. (b) Simulation zoom near resonance comparing free-run and on-resonant injection. Injection produces a sharp coherent peak at the drive frequency atop a suppressed broadband pedestal, consistent with reduction of phase diffusion in the locked state. (c) Measured spectra (peak-centered) for free-running and on-resonant injection (−99.35 dBm injected power). The experimental linewidth collapse and coherent peak formation closely reproduce the simulated entrained spectrum. (d) Gain at the drive frequency versus normalized detuning $\Delta f/f_0$. Left: simulated gain roll-off for multiple injection amplitudes (F = $5\times10^{-1}$ to $10^{-4}$), showing increasing coherent carrier enhancement as detuning → 0. Right: comparison of weak-drive simulation (F = $10^{-4}$) with measurements and a Stuart–Landau phase model fit yielding locking bandwidth $\Gamma = 15.84$kHz. The monotonic detuning dependence confirms injection-mediated enhancement governed by nonlinear amplitude–phase coupling near the Hopf limit cycle.

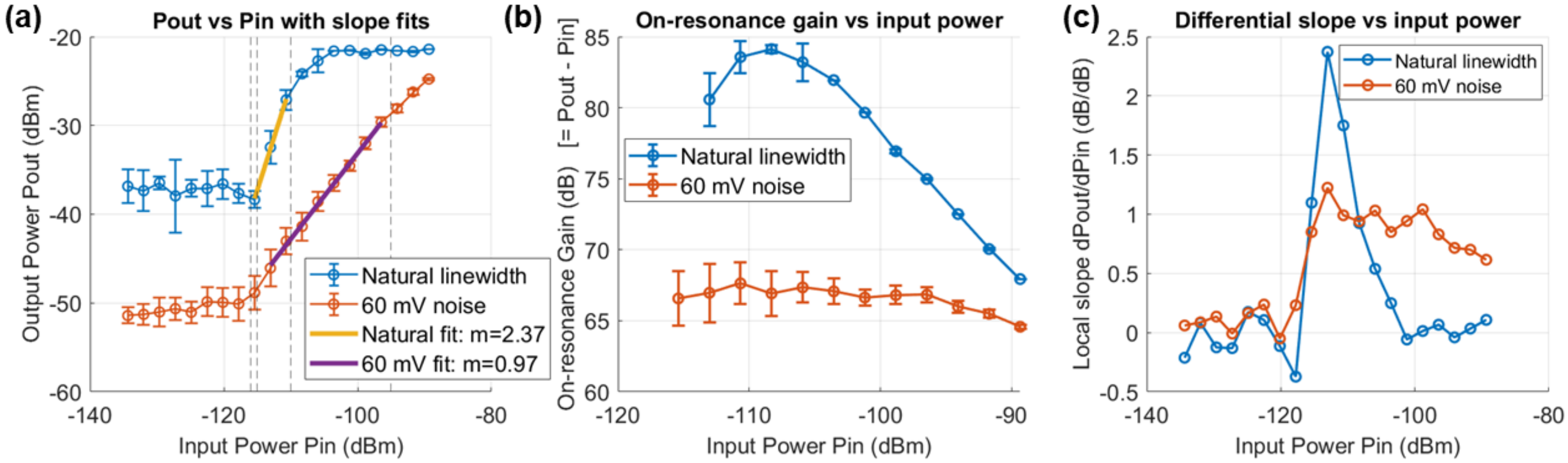

**Fig. 5: Extreme on-resonant reflection gain and its control via phase diffusion.** (a) Coherent reflected carrier power $P_{\text{out}}$ versus injected power $P_{\text{in}}$under on-resonant drive. The natural-linewidth oscillator (blue) exhibits strongly superlinear scaling near the quenching transition with differential slope $m = dP_{\text{out}}/dP_{\text{in}} = 2.37$, indicating nonlinear enhancement of the coherent reflected carrier. Artificially broadening the autonomous linewidth by adding broadband RMS voltage noise (60 mV, red) suppresses the superlinear response ($m = 0.97$). (b) On-resonance reflection gain ($P_{\text{out}} - P_{\text{in}}$ both dB) versus input power calculated from the coherent reflected carrier power in the synchronized regime. With the native oscillator linewidth, the effective reflection gain exceeds 80 dB, corresponding to exceptionally large coherent reflected-carrier enhancement. Linewidth broadening reduces the maximum gain (~67 dB) and flattens the gain–power dependence, consistent with increased phase diffusion. (c) Local differential slope $dP_{\text{out}}/dP_{\text{in}}$extracted from (a). The natural-linewidth oscillator shows a sharp superlinear peak (>2) at the quenching threshold, whereas enhanced phase diffusion broadens and suppresses this peak. These results support competition between injection strength and intrinsic phase diffusion in the coherent reflection response.

To determine how phase diffusion governs the resonant reflection gain, the on-resonant response was measured as a function of injected power under both the natural oscillator linewidth and an artificially broadened linewidth produced by externally applied broadband voltage noise (60mV added noise). Under natural operating conditions, the coherent reflected carrier power at the injected frequency exhibits a strongly superlinear increase near the quenching transition (Fig. 5a), with a differential slope substantially exceeding unity, indicating nonlinear concentration of oscillator energy into the coherent reflected carrier. This quantity represents the power concentrated in the narrow synchronized carrier rather than the total integrated oscillator power. In the free-running state, the autonomous oscillator power is distributed over a broad phase-diffusive spectrum; synchronization collapses the linewidth and concentrates a larger fraction of this power into the coherent carrier. In contrast, artificial linewidth broadening suppresses this superlinear response and produces an approximately linear power scaling, consistent with enhanced stochastic phase diffusion preventing efficient phase localization. The suppression of the coherent response under increased broadband bias noise is consistent with enhanced stochastic phase diffusion limiting phase localization.

The corresponding on-resonant gain, $10\log_{10}(P_{out}/P_{in})$ (in dB), is shown in Fig. 5b. Under natural linewidth conditions, the device reaches peak reflected gain exceeding 80 dB near the quenching threshold, after which the gain gradually decreases as the injected tone increasingly dominates the synchronized response. The reflection-gain metric is applied only after formation of the synchronized carrier and is not interpreted in the extremely weak injection free-running limit (applied for $\gtrsim$-110dBm input power in Fig. 5b). When additional broadband noise is applied, the maximum gain is strongly reduced and the gain curve becomes substantially flatter, demonstrating that excess phase diffusion limits coherent spectral concentration.

This behavior is further emphasized by the local differential slope $dP_{\text{out}}/dP_{\text{in}}$ (both dB) extracted in Fig. 5c. Near the quenching transition, the natural-linewidth oscillator exhibits a pronounced superlinear peak with slope $> 2$, revealing a highly nonlinear synchronization regime in which small increases in injected power produce disproportionately large coherent reflected enhancement. Under enhanced phase diffusion, this peak broadens and collapses toward unity, indicating suppression of the synchronization-driven gain mechanism. Together, these measurements support the interpretation that the large resonant reflection gain is governed by the competition between injection-induced phase localization and phase diffusion of the nonlinear limit cycle. More generally, the observed large reflected enhancement illustrates how synchronization near a noisy Hopf instability can transform a self-oscillatory nonlinear system into an active coherent reflector, enabling large narrowband coherent reflection enhancement through phase ordering.

## Theoretical Framework

### Injection locking of a noisy Hopf oscillator

Near onset, a NDR oscillator can be modeled by the Stuart–Landau (supercritical Hopf) equation[9–11]

$$\dot{z} = (\mu + i\omega_0)z - (1 + i\beta)\,|\,z\,|^2\, z + Fe^{i\omega_{\mathrm{RF}}t} + \eta(t),$$

where $z$ is the complex oscillation envelope, $\mu > 0$ sets the free-running limit-cycle amplitude $r_0 = \sqrt{\mu}$, $\beta$ captures amplitude–phase coupling (nonisochronicity), and $\eta$ represents noise. Here, the Stuart–Landau framework is used as a reduced phenomenological description of the synchronization dynamics rather than as a device-level model of the tunnel-diode circuit. For sufficiently strong radial stability compared with forcing and noise ($2\mu \gg \{D_\phi, |\,F\,|/r_0, |\,\Delta\omega\,|\}$), amplitude is slaved and phase reduction applies[15–17], yielding the noisy Adler equation[18,19]

$$\dot{\psi} = \Delta\omega - \Omega \sin\psi + \xi(t),$$
$$\langle \xi(t)\xi(t') \rangle = 2D_\phi \delta(t - t'),$$

where $\psi = \phi - \omega_{\mathrm{RF}}t$, $\Omega \sim F/r_0$ (renormalized by $\beta$), and the free-running linewidth satisfies $\Gamma \sim D_\phi$[20,21].

### Phase localization and coherent carrier

On resonance ($\Delta\omega = 0$), the stationary phase distribution is the Von Mises form[12,13]

$$p(\psi) = \frac{1}{2\pi I_0(\alpha)} e^{\alpha\cos\psi}, \alpha = \frac{\Omega}{D_\phi},$$

giving the coherence order parameter

$$R = |\,\langle e^{i\psi} \rangle\,| = \frac{I_1(\alpha)}{I_0(\alpha)}.$$

Thus $R \simeq \alpha/2$ for $\alpha \ll 1$ (diffusive regime) and $R \to 1$ for $\alpha \gg 1$ (phase-locked regime). The crossover $\alpha \sim 1$ marks injection-induced quenching of phase diffusion[12,14].

### Reflection gain as frequency concentration

In the rotating frame $u = ze^{-i\omega_{\mathrm{RF}}t} \approx r_0 e^{i\psi}$, the coherent spectral component at $\omega_{\mathrm{RF}}$ scales as

$$|\,\langle u \rangle\,|^2 = r_0^2 R^2.$$

With linear output coupling $\kappa$,

$$P_{\mathrm{out}}(\omega_{\mathrm{RF}}) = \kappa r_0^2 R^2 = \kappa\mu \left[\frac{I_1(\alpha)}{I_0(\alpha)}\right]^2.$$

For weak localization ($\alpha \ll 1$),

$$P_{\mathrm{out}} \propto \frac{F^2}{\Gamma^2},$$

showing linewidth-limited coherent buildup. For strong localization ($\alpha \gg 1$),

$$P_{\mathrm{out}} \to \kappa\mu,$$

so the coherent carrier saturates at the oscillator power set by DC bias and negative resistance[1,2]. Since the injected power required to maintain locking can be small compared with the autonomous oscillation power concentrated into the coherent carrier, the effective narrowband reflection gain can become very large, even though the coherent reflected power remains bounded by the autonomous oscillation power. This effective gain represents the coherent reflected-to-injected carrier power ratio and should not be interpreted as conventional small-signal linear gain.

### Detuning-induced roll-off

Detuning reduces phase localization and thus the coherent carrier. Deterministically, Adler locking exists only within $|\,\Delta\omega\,| < \Omega$, and linearization about the stable fixed point gives an in-well restoring rate[18,19]

$$\lambda(\Delta\omega) \approx \sqrt{\Omega^2 - \Delta\omega^2}.$$

For weak noise, coherence scales with this restoring rate relative to diffusion[12,14], implying rapid suppression of the narrowband gain with detuning and collapse at the unlocking boundary.

## Discussion

The observed large resonant reflection gain originates from a fundamentally nonlinear synchronization process rather than conventional linear amplification. Near the Hopf instability, weak coherent injection suppresses stochastic phase diffusion and localizes the oscillator phase, causing broadband oscillation energy to collapse into a narrow synchronized spectral component. The resulting reflected enhancement therefore does not arise from stimulated gain within a linear susceptibility, but instead from coherent redirection of energy already sustained by the self-oscillatory negative-resistance state. Accordingly, the observed enhancement is interpreted as synchronization-induced coherent spectral concentration rather than as evidence for an increase in total generated RF power. In this regime, the tunnel-diode oscillator behaves as an active nonlinear reflector whose effective reflectivity is dynamically controlled through phase entrainment. The accompanying linewidth collapse, detuning-dependent synchronization bandwidth, and sensitivity to externally induced phase diffusion collectively support phase localization as the mechanism underlying the large narrowband reflected gain.

More broadly, these results establish synchronization-enabled coherent reflection as a general wave-dynamical mechanism that can emerge in noisy self-oscillatory systems near a Hopf bifurcation. While demonstrated here using a tunnel-diode microwave oscillator, the underlying physics is generic to nonlinear limit-cycle systems possessing coherent forcing, intrinsic phase diffusion, and stored oscillation energy. Beyond tunnel diodes, these results suggest a broader class of synchronization-enabled active reflectors in which weak coherent forcing dynamically reorganizes stored oscillatory energy into highly phase-coherent scattered modes, with possible relevance to nonlinear wave control, coupled oscillator networks, and active reconfigurable surfaces.

## Methods

### Experimental setup

The experimental platform consisted of a self-sustained microwave tunnel-diode oscillator (TDO) operated near 2.96 GHz and configured as an active nonlinear reflector through a three-port circulator geometry. The oscillator was constructed using a planar germanium tunnel diode (Eclipse MDI MBD1057-E28, sourced through RFMW) mounted on a custom microwave printed circuit board fabricated on a two-sided Rogers RT/Duroid 6002 substrate. The PCB layout provided RF grounding and DC bias routing for the planar diode package while minimizing parasitic inductance and capacitance associated with the negative differential resistance (NDR) region. The tunnel diode was integrated onto a compact ($\sim 1$ inch) two-sided Rogers RT/Duroid 6002 microwave PCB, where a short microstrip trace biased the diode anode and the cathode was directly connected to the ground plane to minimize parasitic reactances in the NDR oscillation regime.

The tunnel diode was DC biased within its NDR regime to sustain a self-oscillatory microwave limit cycle. RF injection and reflected-signal extraction were implemented using a DiTom Microwave D3C2080 circulator (2–8 GHz operating range), configured such that a weak injected RF tone entered the oscillator through one circulator port while the reflected signal containing both injected and oscillator-generated components was monitored at the output port. A Pasternack PE1639 broadband bias tee (50 kHz–26 GHz) combined the DC bias and RF injection paths. The injected microwave tone was generated using a Signal Hound VSG60A vector signal generator, while the reflected spectra and time-resolved spectrograms were acquired using a Tektronix RSA306B real-time spectrum analyzer (9 kHz–6.2 GHz bandwidth).

The free-running oscillation frequency was approximately 2.962 GHz, with small shifts depending on DC bias and operating conditions. Measurements of injection pulling and injection locking were performed by sweeping either injected RF power or injection frequency detuning relative to the autonomous oscillator frequency. For injection-pulling measurements (Fig. 3a), the injected power was varied from approximately $-134$dBm to $-89$dBm while recording real-time reflection spectrograms. Injection-locking measurements (Fig. 3b) were obtained by sweeping the injected RF frequency across the free-running oscillator resonance over detuning ranges up to several hundred kilohertz. The locking bandwidth was extracted from the detuning interval over which the oscillator frequency remained entrained to the injected tone.

Reflection gain is defined as $G = 10\log_{10}(P_{out}/P_{in})$ (in dB), where $P_{out}$ is the coherent reflected power at the synchronized carrier frequency and $P_{in}$ is the injected RF power referenced to the tunnel-diode input plane after accounting for transmission-path, bias-tee, and circulator insertion losses. This metric is evaluated only in the synchronized regime and represents the coherent reflected-to-injected carrier power ratio rather than conventional small-signal amplifier gain. Measurements with the tunnel diode off were used to establish the direct injected-carrier and measurement-chain feedthrough baseline in the absence of the self-sustained oscillation. Instantaneous linewidths were estimated from the measured spectral full-width at half-maximum (FWHM) of the coherent reflected peak obtained from the real-time spectral measurements.

Time–frequency spectrograms were generated directly from continuous real-time spectrum analyzer acquisitions during power or frequency sweeps.

To investigate the role of phase diffusion, controlled broadband voltage noise was added to the DC bias line using a BK Precision 4064B dual-channel function generator operated in broadband noise mode. RMS noise amplitudes up to 60 mV were injected through the bias circuitry, intentionally broadening the autonomous oscillator linewidth and increasing stochastic phase diffusion of the limit cycle. Comparative measurements with and without externally applied noise were used to quantify the sensitivity of the synchronization-induced reflection gain to phase diffusion and linewidth broadening.

## Data availability

Measurement results in Figs. 1 and 3, including linewidths, injection vs. reflected power, frequency-detuning sweeps, and spectrogram-derived data traces, are available as source data. All other data is available upon request.

**Figure Legends:**

**Fig. 1: Injection-induced quenching concentrates oscillator power into a coherent reflected carrier:** Increasing resonant injection drives a transition from a free-running noisy oscillation to a phase-localized state in a tunnel-diode oscillator (TDO). As the injected power increases, the instantaneous linewidth (right axis) collapses while the coherent reflected power (left axis) rises sharply, revealing suppression of phase diffusion and concentration of oscillation energy into a synchronized spectral component. The dashed line marks the onset of quenching, where coherent ordering emerges and reflected power rapidly increases. The inset shows representative spectra illustrating the transformation from a broadband free-running state (FR) to a narrow-quenched state (Q).

**Fig. 2: On-resonant reflection gain in a tunnel-diode oscillator (TDO).** (a) Tunnel-diode I–V characteristic highlighting the negative differential resistance (NDR, dI/dV < 0) region that provides negative damping and sustains a nonlinear microwave limit cycle. (b) Experimental configuration: a DC-biased tunnel diode (TD) coupled via a circulator; a weak injected RF tone drives the TDO and the reflected port contains both injected and oscillator fields. (c) Nonlinear reflector concept: the self-oscillatory (OSC) diode acts as an active mirror whose reflection is governed by Hopf-type amplitude–phase dynamics. (d) Measured reflection spectra near 2.962 GHz for varying injection detuning frequency, showing progressive enhancement of the coherent reflected carrier as the detuning approaches zero. (e) Time–frequency spectrogram during a frequency sweep across OSC resonance; as detuning approaches zero, the OSC linewidth collapses and power concentrates into the synchronized component (quenching). (f) Reflection gain versus detuning; data follow a Stuart–Landau roll-off with locking bandwidth Γ = 15.84 kHz, consistent with injection-modified Hopf dynamics. (g) Residual phase portraits in the rotating frame ($u - \langle u \rangle$); F is normalized injection amplitude and $R = |\langle e^{i\phi} \rangle|$ the phase-coherence order parameter. Increasing F (0→0.5) drives R≈0.15→0.99 and shrinks the residual cloud toward the coherent fixed point (origin). (h) Measured spectrogram identifying the free-running oscillator (OSC), injection-locked (Locked), and tunnel-diode-off (TD Off) conditions, with the TD Off condition providing the direct injected-carrier/feedthrough baseline. (i) Representative reflection spectrum under strong injection showing OSC and reflected signal relative to injected RF (INJ) and system noise floor.

**Fig. 3: Injection pulling and locking dynamics of the tunnel-diode oscillator.** (a) Injection pulling (power sweep). Measured reflection spectrogram near 2.9GHz while sweeping injected RF power from −134dBm to −89dBm. The horizontal branch corresponds to the free-running self-sustained oscillation (OSC). Increasing injection power

pulls the oscillator frequency toward the injected tone (INJ RF) through nonlinear amplitude–phase coupling of the NDR-driven limit cycle. The strong curvature at highest powers marks approach to synchronization. (b) Injection locking (frequency sweep). Reflection spectrogram versus injected-frequency detuning (−0.4 to +0.6 MHz) at fixed injection amplitude. The diagonal trace denotes the swept INJ RF. Within a finite detuning interval (dashed lines), the oscillator is frequency-entrained (IL), collapsing onto the drive. Outside this interval, injection unlocking (IUL) occurs and the oscillator reverts to detuned free-running behavior. The locking bandwidth is defined by the detuning span over which entrainment persists, consistent with Adler/Stuart–Landau phase dynamics.

**Fig. 4: Noisy Hopf model of injection-enhanced reflection and detuning-dependent gain.** (a) Ensemble-averaged power spectra of a noisy Hopf oscillator (μ = 0.10, σ = 0.10, F = 0.05), normalized to the free-running peak and re-centered to the measured natural frequency $f_0$. Free-run (F = 0), on-resonance injection, and positive detunings (+2–15%) are shown. As detuning decreases, spectral weight collapses into the injected component and the linewidth narrows, evidencing phase entrainment and quenching of incoherent fluctuations. (b) Simulation zoom near resonance comparing free-run and on-resonant injection. Injection produces a sharp coherent peak at the drive frequency atop a suppressed broadband pedestal, consistent with reduction of phase diffusion in the locked state. (c) Measured spectra (peak-centered) for free-running and on-resonant injection (−99.35 dBm injected power). The experimental linewidth collapse and coherent peak formation closely reproduce the simulated entrained spectrum. (d) Gain at the drive frequency versus normalized detuning $\Delta f/f_0$. Left: simulated gain roll-off for multiple injection amplitudes (F = $5\times10^{-1}$ to $10^{-4}$), showing increasing coherent carrier enhancement as detuning → 0. Right: comparison of weak-drive simulation (F = $10^{-4}$) with measurements and a Stuart–Landau phase model fit yielding locking bandwidth $\Gamma = 15.84$kHz. The monotonic detuning dependence confirms injection-mediated enhancement governed by nonlinear amplitude–phase coupling near the Hopf limit cycle.

**Fig. 5: Extreme on-resonant reflection gain and its control via phase diffusion.** (a) Coherent reflected carrier power $P_{\text{out}}$ versus injected power $P_{\text{in}}$under on-resonant drive. The natural-linewidth oscillator (blue) exhibits strongly superlinear scaling near the quenching transition with differential slope $m = dP_{\text{out}}/dP_{\text{in}} = 2.37$, indicating nonlinear enhancement of the coherent reflected carrier. Artificially broadening the autonomous linewidth by adding broadband RMS voltage noise (60 mV, red) suppresses the superlinear response ($m = 0.97$). (b) On-resonance reflection gain ($P_{\text{out}} - P_{\text{in}}$ in dB) versus input power calculated from the coherent reflected carrier power in the synchronized regime. With the native oscillator linewidth, the effective reflection gain exceeds 80 dB, corresponding to exceptionally large coherent reflected-carrier enhancement. Linewidth broadening reduces the maximum gain (~67 dB) and flattens the gain–power dependence, consistent with increased phase diffusion. (c) Local differential slope $dP_{\text{out}}/dP_{\text{in}}$extracted from (a). The natural-linewidth oscillator shows a sharp superlinear peak (>2) at the quenching threshold, whereas enhanced phase diffusion broadens and suppresses this peak. These results support competition between injection strength and intrinsic phase diffusion in the coherent reflection response.

## Acknowledgements

The author would like to acknowledge discussions with A. Ferraz at JPL (Jet Propulsion Laboratory, California Institute of Technology), and G. Durgin at Georgia Tech (Georgia Institute of Technology). The research was carried out at the Jet Propulsion Laboratory, California Institute of Technology, under a contract with the National Aeronautics and Space Administration (80NM0018D0004), through the Satellite Needs Working Group (SNWG).

## Author contributions

D.A. conceived the study, developed the theoretical framework, designed and performed the experiments, carried out the measurements and data analysis, developed the numerical simulations, interpreted the results, and wrote the manuscript. B.F. designed and fabricated the microwave tunnel-diode PCB assembly and supported initial experimental testing of the tunnel-diode oscillator platform. J.B. supported programmatic coordination, experimental data collection efforts, and literature review. All authors discussed the results and reviewed the manuscript.

## Additional information

The author declares no competing interest.

## Supplementary information

## Supplementary Section 1

### Linewidth Estimation and Robustness of the Quenching Transition

A central signature of the injection-induced quenching transition is the abrupt collapse of the oscillator spectral linewidth as the injected signal approaches the synchronization threshold. Because linewidth extraction can depend on the specific definition used to quantify spectral width, an additional analysis was performed to verify that the observed linewidth suppression is not an artifact of the chosen linewidth metric. Supplementary Fig. 1 shows the estimated linewidth as a function of injected power using multiple peak-drop definitions corresponding to $-3$, $-10$, $-15$, and $-20$dB thresholds relative to the instantaneous spectral maximum.

The linewidths were extracted directly from time-resolved real-time spectrum analyzer acquisitions recorded during injected-power sweeps. For each injected-power dwell interval, the oscillator spectrum was first peak-tracked within a local search window surrounding the carrier frequency. The spectral width was then computed by identifying the frequency crossings at fixed power thresholds below the instantaneous peak amplitude. Linear interpolation between neighboring spectral bins was used to estimate the threshold-crossing frequencies with sub-bin resolution. This procedure was independently repeated for each threshold level and for all spectra acquired within a given dwell interval, after which the linewidth mean and standard deviation were computed across the ensemble of measured spectra.

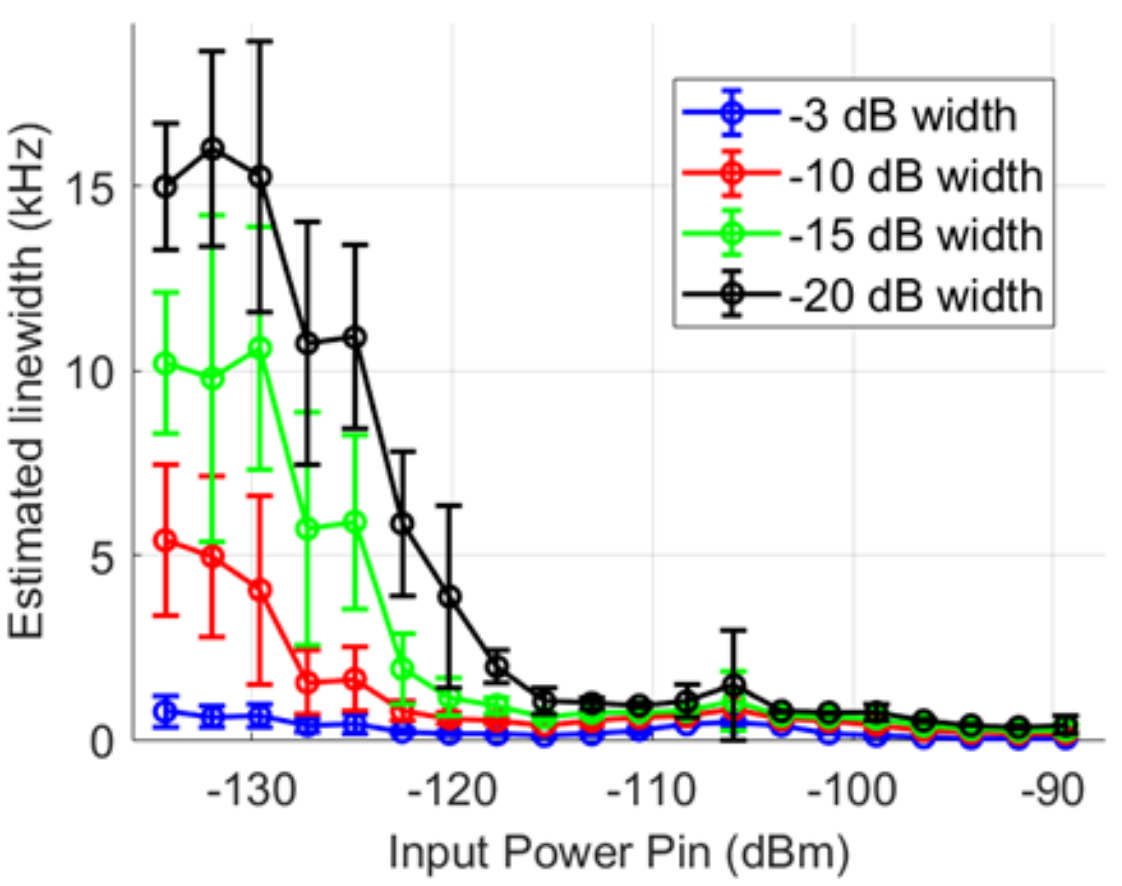

**Supplementary Fig. 1: Linewidth collapse.** Estimated spectral linewidth as a function of injected power for several linewidth definitions based on different peak-drop thresholds (−3, −10, −15, and −20 dB relative to the spectral maximum). Although the absolute linewidth values vary with the chosen threshold, all definitions exhibit the same qualitative behavior: a rapid collapse of linewidth near the quenching onset followed by a narrow, stabilized spectral peak at higher injection levels. This demonstrates that the observed linewidth suppression reported in Fig. 1 of main manuscript is not sensitive to the linewidth metric used to quantify the spectral width.

As shown in Supplementary Fig. 1, all linewidth definitions exhibit the same qualitative behavior despite substantial differences in absolute linewidth values. In the weakly injected regime, the oscillator exhibits a broad phase-diffusive spectrum whose apparent width depends strongly on the chosen threshold because of the extended incoherent pedestal. As the injected power approaches the quenching threshold, all linewidth estimates collapse rapidly and converge toward a narrow stabilized spectral feature. At higher injection levels, the linewidth remains narrow and largely independent of threshold definition, indicating formation of a strongly phase-localized synchronized carrier. The agreement among the different linewidth metrics demonstrates that the linewidth suppression reported in the main text is a robust manifestation of synchronization-induced quenching rather than a consequence of the specific linewidth extraction procedure.

## Supplementary Section 2

### Integrated Channel Power and Spectral Concentration

While the linewidth collapse directly reveals suppression of phase diffusion, it is also important to quantify how the oscillator power redistributes spectrally during synchronization. To examine this behavior, the integrated spectral power surrounding the oscillator carrier was analyzed as a function of both injection strength and analysis bandwidth. Supplementary Fig. 2 shows the integrated channel power versus analysis bandwidth for several representative injected-power conditions spanning the free-running, transitional, and strongly synchronized regimes.

The integrated channel power was computed directly from the measured spectrogram data by first identifying the instantaneous spectral peak for each acquired spectrum and then integrating the spectral power within a symmetric frequency window centered on the tracked carrier. Multiple integration bandwidths ranging from sub-kilohertz to tens of kilohertz were evaluated. The spectral power contained within each window was converted from dBm to linear power units prior to integration and then converted back into dBm for reporting. The integration procedure was repeated across all spectra acquired during each injected-power dwell interval, allowing the mean integrated channel power and corresponding statistical variation to be determined for each bandwidth and injection level.

In the free-running regime ($P_{\text{in}} = -134.3$ dBm), the oscillator exhibits a broad incoherent spectral distribution, causing the recovered channel power to increase strongly with integration bandwidth as progressively more of the phase-diffusive pedestal is included. As the injected power increases ($-127.2$ and $-120.1$dBm), the oscillator spectrum narrows and a larger fraction of the oscillation energy becomes concentrated near the carrier frequency, reducing the bandwidth dependence of the recovered power. In the strongly synchronized regime ($P_{\text{in}} = -98.8$ dBm), the integrated power becomes nearly independent of analysis bandwidth, indicating that most of the oscillator energy has collapsed into a narrow coherent spectral component. This behavior provides additional evidence that injection-induced quenching acts by concentrating previously broadband oscillation energy into a highly phase-coherent synchronized carrier.

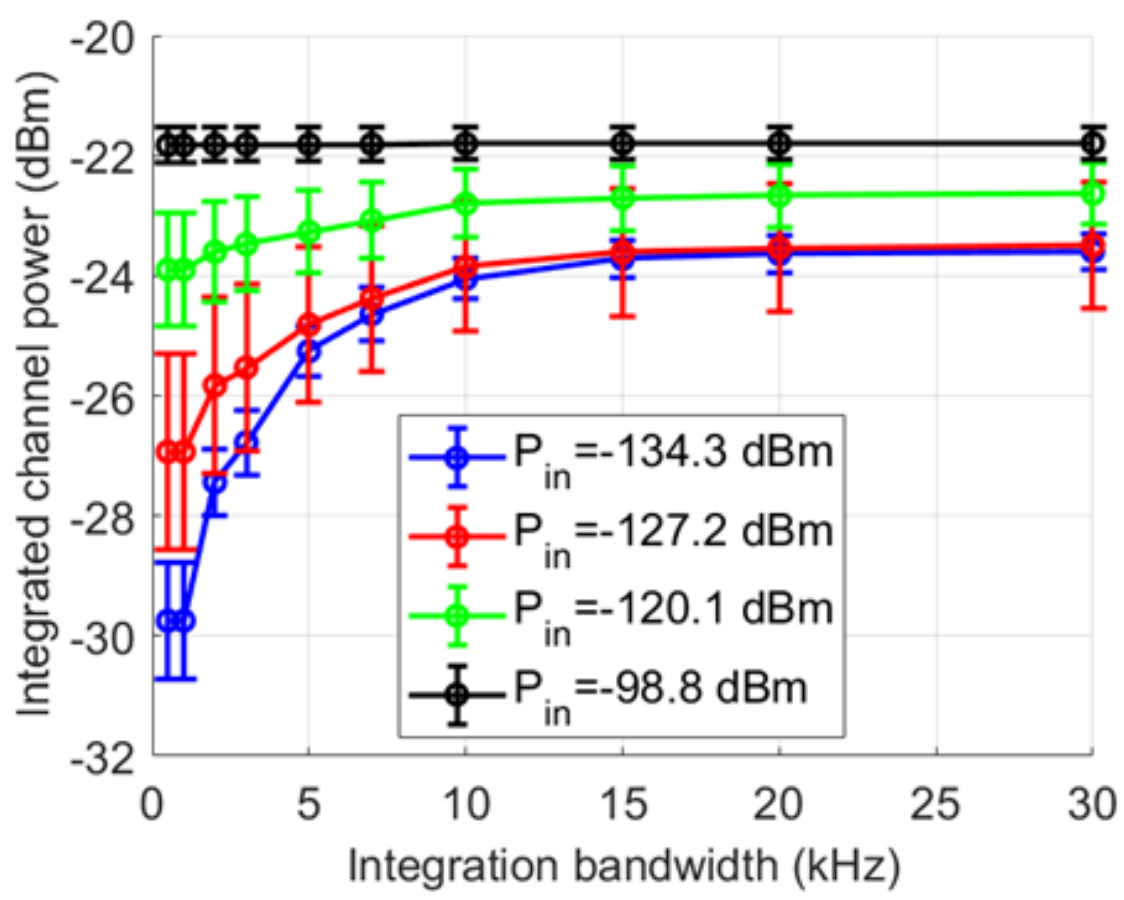

**Supplementary Fig. 2: Integrated channel power versus analysis bandwidth for different injection levels.** The spectral power integrated around the oscillator carrier is plotted as a function of analysis bandwidth for several injected powers $P_{in}$. In the free-running regime ($P_{in} = -134.3$ dBm), the oscillator exhibits a broad spectrum, and the recovered channel power increases rapidly with bandwidth as more of the spectral pedestal is included. As the injected signal increases ($-127.2$ and $-120.1$ dBm), the spectrum narrows and a larger fraction of the oscillator power is concentrated near the carrier frequency, causing the integrated power to rise more slowly. In the strongly injected regime ($P_{in} = -98.8$ dBm), most of the oscillator energy resides in a narrow coherent carrier and the integrated power becomes nearly independent of bandwidth. This behavior illustrates the progressive spectral concentration of oscillator power accompanying injection-induced quenching.